\documentclass[conference]{IEEEtran}
\IEEEoverridecommandlockouts
\usepackage{amsmath,amssymb,amsfonts}
\usepackage{algorithmic}
\usepackage{graphicx}
\usepackage{textcomp}
\usepackage{float}
\usepackage{listings}
\usepackage{xcolor}
\usepackage{array}
\usepackage{longtable}
\usepackage{adjustbox}
\usepackage{multirow}
\usepackage{rotating}
\usepackage[backend=biber,bibencoding=latin1, sorting=none]{biblatex}
\def\BibTeX{{\rm B\kern-.05em{\sc i\kern-.025em b}\kern-.08em
    T\kern-.1667em\lower.7ex\hbox{E}\kern-.125emX}}
\begin{document}

\lstset
{ %Formatting for code in appendix
    language=Java,
    basicstyle       = \ttfamily,
    commentstyle=\bold,
    aboveskip=0mm,
    belowskip=0mm,
    showstringspaces=false,
    columns=flexible,
    basicstyle={\fontsize{8}{7}\selectfont\ttfamily},
    numbers=left,
    numberstyle=\tiny\color{gray},
    keywordstyle=\color{blue},
    commentstyle=\color{ao},
    stringstyle=\color{mauve},
    breaklines=true,
    breakatwhitespace=true,
    tabsize=3,
    escapeinside={<@}{@>},
    xleftmargin=0.5cm
}

\title{Impact of Code Transformation on Detection of Smart Contract Vulnerabilities\\
}

\author{\IEEEauthorblockN{Cuong Tran Manh}
\IEEEauthorblockA{\textit{Faculty of Information Technology} \\
\textit{VNU University of Engineering  }\\ \textit{and Technology} \\
Hanoi, Vietnam \\
tranmanhcuong@vnu.edu.vn}
\and
% \IEEEauthorblockN{Thu-Trang Nguyen}
% \IEEEauthorblockA{\textit{Faculty of Information Technology} \\
% \textit{VNU University of Engineering  }\\ \textit{and Technology} \\
% Hanoi, Vietnam \\
% trang.nguyen@vnu.edu.vn}
% \and
\IEEEauthorblockN{Hieu Dinh Vo}
\IEEEauthorblockA{\textit{Faculty of Information Technology} \\
\textit{VNU University of Engineering  }\\ \textit{and Technology} \\
Hanoi, Vietnam \\
hieuvd@vnu.edu.vn}
}

\maketitle

\begin{abstract}
While smart contracts are foundational elements of blockchain applications, their inherent susceptibility to security vulnerabilities poses a significant challenge. Existing training datasets employed for vulnerability detection tools may be limited, potentially compromising their efficacy. This paper presents a method for improving the quantity and quality of smart contract vulnerability datasets and evaluates current detection methods. The approach centers around semantic-preserving code transformation, a technique that modifies the source code structure without altering its semantic meaning. The transformed code snippets are inserted into all potential locations within benign smart contract code, creating new vulnerable contract versions. This method aims to generate a wider variety of vulnerable codes, including those that can bypass detection by current analysis tools. The paper experiments evaluate the method's effectiveness using tools like Slither, Mythril, and CrossFuzz, focusing on metrics like the number of generated vulnerable samples and the false negative rate in detecting these vulnerabilities. The improved results show that many newly created vulnerabilities can bypass tools and the false reporting rate goes up to 100\% and increases dataset size minimum by 2.5X.
\end{abstract}


\section{Introduction}
The rapid proliferation of blockchain technology has spurred its exploration across a multitude of domains, encompassing finance, supply chain management, and value chain applications \cite{yaga2019blockchain, colin2024integrated}. However, alongside these promising advancements, inherent security risks associated with blockchain technology necessitate careful consideration \cite{dhawan2017analyzing}. Vulnerabilities in smart contracts are one of the most serious threats, in the 2016 DAO attack, the Reentrancy vulnerability was exploited causing a loss of 60 million US dollars. This incident underscores the urgent need for robust security measures to safeguard smart contracts, especially considering the vast sums of money they can potentially manage, often reaching tens of billions of US dollars \cite{liu2021combining}. 

Recently, many commercial and research tools have focused on detecting vulnerabilities in smart contracts, and new techniques such as applying machine learning and deep learning are also gradually being used \cite{liu2021combining, mi2021vscl, kien2023multimodal, sun2023assbert, zhen2024gnn}. Almost all tools have a certain false positive/negative rate \cite{ghaleb2020effective}. 

Moreover, the approaches that use deep learning require a lot of data. 
Unfortunately, currently, the number of quality datasets is limited. All human-verified datasets are under 1000 vulnerability source codes \cite{chu2023survey}. Meanwhile, datasets with more vulnerability samples are only verified by static analysis tools, which is less reliable. For example, the SoliAudit dataset \cite{liao2019soliaudit} consists of about 20 thousand smart contracts, which have been used to evaluate many recent research methods. However, this dataset uses tools for validation, which raises concerns about not detecting false results missed by these validation tools. Another widely used dataset is SolidiFI, which takes a different approach by using available vulnerability templates and then injecting these vulnerabilities into normal code \cite{ghaleb2020effective} to create source code files containing software vulnerabilities. SolidiFI inserts vulnerabilities at all potential locations and is also considered helpful for finding corner cases that are difficult to detect in real-life scenarios \cite{ghaleb2022towards, colin2024integrated}. This method and dataset are highly reliable when these vulnerabilities have been verified. However, this dataset includes over 9,000 vulnerabilities, which is relatively small given that they represent only 7 distinct vulnerability types, with fewer than 50 samples per type.

Because of the aforementioned situations, this paper aims to enhance the quantity of vulnerability datasets for smart contracts to mitigate potential associated risks. Additionally, to enhance the quality, the dataset needs to have the capability to identify numerous cases of false reports that current state-of-the-art analysis tools may miss. Specifically, we propose an approach using \textit{semantic-preserving code transformation} to create a new vulnerability template while retaining its original attributes. After that, these samples are inserted into potential locations within the source code to generate a contract containing the vulnerability.

To assess the effectiveness of our proposed approach, we conducted a comprehensive evaluation centered on two key aspects: the quantity of generated vulnerability snippets and their qualitative across various analysis tools. Specifically, we designed and executed experiments to analyze the false negative rates exhibited by prominent tools like Slither, Mythril, and CrossFuzz when evaluated against the new dataset.

In summary, this paper aims to address the two questions:
\begin{itemize}
\item \textbf{RQ1:} How does code transformation affect tools that detect vulnerabilities? To find out, we must assess the quantity and quality of vulnerable examples within the newly generated dataset.
\item  \textbf{RQ2:} Is the injection method at all potential locations in the previous research approaches more effective than just one location?
\end{itemize}

\section{Motivation}
Semantic-preserving code transformation is a technique for modifying the structure of source code while ensuring the program's functional behavior remains unaltered. For example, refactoring operations that rename variables exemplifies this principle, as program execution remains unaffected despite the change in variable names. Furthermore, leveraging the programming language's specification, we can exploit alternative syntactic constructs that offer equivalent functionality, such as \texttt{for} and \texttt{while} loops. By adhering to this concept, it becomes possible to generate numerous variations of a source code base while preserving its underlying logic and intended outcomes. Consequently, a source code harboring vulnerabilities will continue to exhibit those vulnerabilities even after transforming, and conversely, a vulnerability-free source code will remain secure.

\begin{figure}[H]
    \centering
    \lstset{
        lineskip={1.85pt}
    }
\begin{lstlisting}[language=Java]
result = DOWN;
if(finalPrice > initialPrice){
    result = UP;
}
\end{lstlisting}
    \caption{Simple original solidity source code}
    \label{fig:example1}
\end{figure}{}

\begin{figure}[H]
    \centering
    \lstset{
        lineskip={1.85pt}
    }
\begin{lstlisting}[language=Java]
result = DOWN;
for(uint i = 0; finalPrice > initialPrice; i++){
    result = UP;
    break;
}
\end{lstlisting}
    \caption{Modified version of Fig.~\ref{fig:example1}}
    \label{fig:example1_mo}
\end{figure}{}

\begin{figure}[H]
    \centering
    \includegraphics[height=155pt]{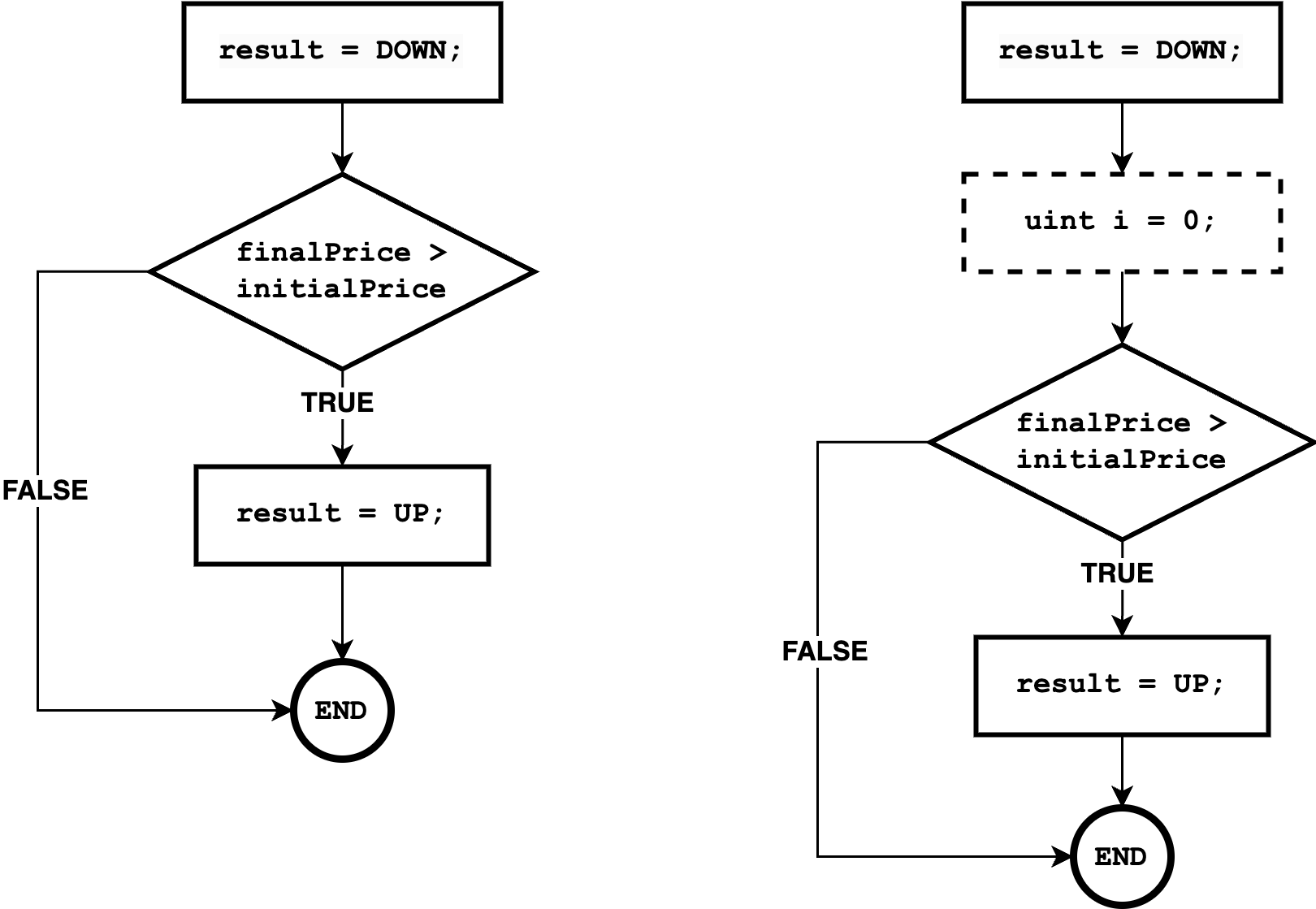}
    \caption{CFG of two programs: Fig.~\ref{fig:example1} and Fig.~\ref{fig:example1_mo}} 
    \label{fig:cfg_smae}
\end{figure}

Two code snippets, illustrated in Fig.~\ref{fig:example1} and~\ref{fig:example1_mo}, demonstrate the concept of semantic equivalence despite syntactic variations. Although both programs achieve the same outcome, they employ distinct constructs and syntaxes. Fig.~\ref{fig:example1} utilizes an if statement on Line 2 for conditional execution, while Fig.~\ref{fig:example1_mo} leverages a for loop. However, the for loop in Fig.~\ref{fig:example1_mo} iterates only once due to the presence of a break statement after the first iteration, effectively mimicking the behavior of an if statement. This semantic equivalence is further corroborated by the Control-Flow Graph (CFG) depicted in Fig.~\ref{fig:cfg_smae}. Disregarding the \texttt{uint i = 0;} statement, which has no meaning to the program's logic, both code snippets exhibit identical control flow. It's noteworthy that the \texttt{break;} statement serves as a control flow construct, dictating program execution paths, while the \texttt{i++} statement within the loop becomes dead code as it's never executed.

The variations of source code pose a significant challenge for vulnerability detection methods. Due to the existence of numerous code structures that can achieve the same program outcome or mean semantic equivalence, traditional analysis techniques may struggle to identify vulnerabilities across these variations \cite{zhang2023challenging}. This variability necessitates addressing two fundamental questions:
\begin{itemize}
    \item Capability of existing methods: Can current vulnerability detection methods effectively transcend syntactic variations and pinpoint vulnerabilities even if they are expressed in a different code structure with equivalent semantics?
    \item Generating variations for improvement: How can we strategically generate additional code variations while preserving the program's core logic? This ability to create diverse code sets would enable the development of more robust detection methods, equipping them to handle the broader spectrum of code structures that may harbor vulnerabilities that are not detected.
\end{itemize}

\section{The Proposed Method}
\begin{figure}[H]
    \centering
    \includegraphics[width=0.485\textwidth]{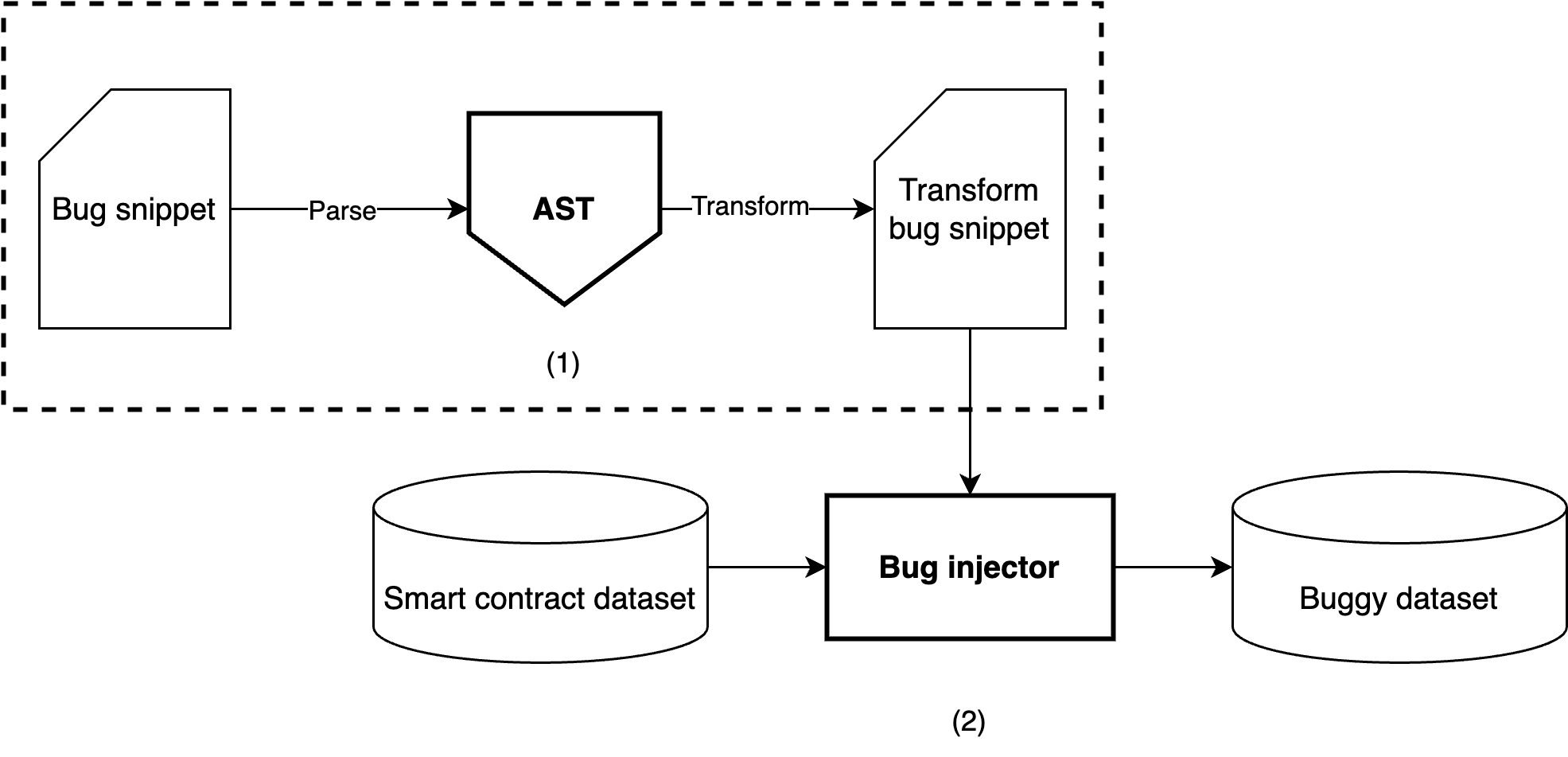}
    \caption{The proposed method workflow} 
    \label{fig:workflow}
\end{figure}

\bgroup
\def\arraystretch{1.35}
\begin{table*}[]
\caption{All transform operators in this method}
\label{tab:operators}
\centering
\begin{tabular}{|>{\hspace{0pt}}m{0.2\linewidth}|>{\hspace{0pt}}m{0.1\linewidth}|>{\hspace{0pt}}m{0.48\linewidth}|>{\hspace{0pt}}m{0.12\linewidth}|} 
\hline
\multicolumn{1}{|c|}{\textbf{Operator}} &
  \multicolumn{1}{c|}{\textbf{Formula}} &
  \multicolumn{1}{c|}{\textbf{Describe}} &
  \textbf{Level} \\ \hline
Variable renaming & \multirow{2}{*}{$Op_{rename}$} & Rename the variable declared in the vulnerability snippet to a different variable name. & \multirow{2}{*}{Naming-level} \\ \cline{1-1} \cline{3-3}
Function renaming &  & Rename the function declared in the vulnerability snippet to a different function name. &  \\ \hline
Permutation                                & $Op_{permutation}$ 
$Op_{subtraction}$ 
$Op_{devision}$ 
$Op_{unequal}$ 

& Swap the position of operands in a binary expression.                      & Expression-level \\ \hline
If branch swapping &
  $Op_{if}$ &
  Invert the if statement condition and swap the position of both true/false statements. &
  \multirow{5}{*}{Statement-level} \\ \cline{1-3}
If statement to for statement transforming   & $Op_{if2for}$      & Replace the if statement with a for loop.                                  &                  \\ \cline{1-3}
If statement to while statement transforming & $Op_{if2while}$    & Replace the if statement with a while loop.                                &                  \\ \cline{1-3}
Variable passing                           & $Op_{tx}$          & Declare and use a new variable instead \texttt{tx.origin} expression. &                  \\ \hline
\end{tabular}
\end{table*}
\egroup

The process depicted in Fig.~\ref{fig:workflow} consists of two main steps: (1) \textit{transforming} the source code containing the vulnerability and (2) \textit{injecting} the vulnerability snippet into the source code of the smart contract. In the first step, the source code containing the vulnerability is converted into an intermediate structure, Abstract Syntax Tree (AST), and then this AST is processed by a transform operator while ensuring the original semantics are preserved. The result of this transformation is a new source code snippet, which still contains vulnerabilities nature. Next, in the second step, the source code snippet containing these vulnerabilities is injected into all potential locations in a smart contract source code, to create new source code versions that have a vulnerability. This also ensures that the experimental results are not affected by the injection location. These source code versions are then syntactically checked and compiled testing to ensure validity. Each location where a vulnerability is added without causing a compilation error is considered a valid vulnerability called by \textit{vulnerable location}.

The transform operators were chosen based on similar studies on semantic-preserving code transformation approaches \cite{zhang2023challenging, le2024evaluating} and combined with the Solidity language specification. The operators were classified into three distinct levels based on the scope of their changing: naming-level, expression-level, and statement-level \cite{le2024evaluating}. We provide a total of 7 specific stand-alone transform operators shown in Table~\ref{tab:operators}.

\subsection{Naming-level}
At this level, two operators are introduced: variable renaming and function renaming. The variable renaming operator systematically identifies and assigns new, valid Solidity identifiers to all declared variables within the code snippet, encompassing both initial declarations and subsequent uses. Importantly, this renaming is confined to the snippet to avoid unintended impact on the main program. Similarly, the function renaming operator identifies and renames declared functions and their calls within the snippet, adhering to Solidity's naming conventions while maintaining a restricted scope to ensure program integrity. These operators pave the way for generating diverse code variations with altered structures but equivalent functionality within the original snippets.
\subsection{Expression-level}
Permutation is employed as a code transformation technique to assess analysis tools' capability in comprehending an expression's value. Permutation is used for binary expression operations. This focuses on exchanging the positions of operands in summation, multiplication, equality comparison, and slightly more complex subtraction, division, and unequal, as illustrated by the following equations~\ref{operator:p1}, \ref{operator:p2}, \ref{operator:p3} and \ref{operator:p4}:
\begin{itemize}
    \item Summation, multiplication, and equality comparison, this operator only swaps the positions of the two operands:
    \begin{equation}
        \label{operator:p1}
        Op_{permutation} = \frac{E_{1} \rightarrow E_{2}  ~~ E_{2} \rightarrow E_{1}}{E_{1}~op~E_{2} \rightarrow E_{2}~op~E_{1}}
    \end{equation}

    \item For subtraction, in addition to reversing the two operands, add the - operator before the expression:
    \begin{equation}
        \label{operator:p2}
        Op_{subtraction} = \frac{E_{1} \rightarrow E_{2}  ~~ E_{2} \rightarrow E_{1}}{E_{1}~op~E_{2} \rightarrow -(E_{2}~op~E_{1})}
    \end{equation}
    \item For division, in addition to inverting the two operands, add the numerator 1/ and the inversion expression to the new denominator:
    \begin{equation}
        \label{operator:p3}
        Op_{division} = \frac{E_{1} \rightarrow E_{2}  ~~ E_{2} \rightarrow E_{1}}{E_{1}~op~E_{2} \rightarrow 1/(E_{2}~op~E_{1})}
    \end{equation}
    \item For unequal comparison operations, add a negation symbol after the permutation:
    \begin{equation}
        \label{operator:p4}
        Op_{unequal} = \frac{E_{1} \rightarrow E_{2}  ~~ E_{2} \rightarrow E_{1}}{E_{1}~op~E_{2} \rightarrow \neg(E_{2}~op~E_{1})}
    \end{equation}
\end{itemize}

Symbols in this paper: $\rightarrow$ means transformation, $E$ is an expression and $S$ is a statement.

\subsection{Statement-level}
This level has four operators: if branch swapping, if statement to for/while statement transforming, and variable passing.

\textit{\textbf{If branch swaping}}. This operator will reverse the position of both sides of the if statement, and also reverse the value of the conditional expression. Two transformations are performed simultaneously, and the flow of the if statement will still be preserved. With this operator, the goal will be to check whether the tools understand the negative expression or not and evaluate whether the position of the conditional branch has any effect on the report result. Formula~\ref{op:if} represents this operator.
\begin{equation}
\label{op:if}
Op_{if} = \frac{E  ~~ S_{T} \rightarrow S_{F} ~~ S_{F} \rightarrow S_{T}}{if(E) S_{T} ~else~ S_{F} \rightarrow if(\neg E) S_{F} ~else~ S_{T}}
\end{equation}

\textit{\textbf{If statement to for/while statement transforming}}. The essence of a loop is to check the condition before or after each iteration. Therefore, if the loop statement is executed only once, this is equivalent to checking the condition only once, similar to an if statement. However, the loop structure is much more complicated than the two branches of an if conditional. The goal of converting an if statement to a for or while loop is to make the control flow of the program more complicated while adding several statements that do not affect the meaning of the program. This can make it difficult for analysis tools not to handle the above loop structures strongly enough, or to understand how many loops there are. This operator only performs transform on conditional statements with only one branch. Two operator are represented by equations~\ref{op:if2while} and \ref{operator:if2for}.

\begin{equation}
\label{op:if2while}
    Op_{if2while}=\frac{E ~ S}{if(E) ~ S \rightarrow while(E) \{S;break;\}}
\end{equation}

\begin{equation}
    \label{operator:if2for}
    Op_{if2for}=\frac{E ~ S}{if(E) ~ S \rightarrow for(s_{1};E;s_{2}) \{S;break;\}}
\end{equation}

Equation~\ref{operator:if2for} $s_{1}$ includes $s_{1}$ and $s_{2}$ as two optional statements. They have no semantic meaning in the program like the example in Fig.~\ref{fig:example1_mo}. At the same time, these statements are more complex and can affect gas consumption. According to the Solidity specification, we can use two empty statements at this location or declare a temporary variable.

\textbf{\textit{Variable passing.}} This operator targets the \texttt{tx.origin} expression within Solidity smart contracts. The \texttt{tx.origin} variable holds the address of the original transaction initiator, which can be exploited by malicious contracts for deceptive purposes \cite{kado2023empirical}. This operator will assign a variable using \texttt{tx.origin} to avoid using this expression directly. The analysis phase potentially ignores this vulnerability when there is no direct usage statement.

\begin{equation}
    \label{operator:tx}
    Op_{tx} = \frac{S(tx.origin) \rightarrow S'(v)}{S(tx.origin) \rightarrow v = tx.origin;S'(v)}
\end{equation}

Variable passing operator for \texttt{tx.origin} is represented by Formula~\ref{operator:tx} with $v$ as an added intermediate variable.

\begin{figure}[H]
    \centering
    \lstset{
        lineskip={1.85pt}
    }
\begin{lstlisting}[language=Java]
{
	require (tx.origin == owner_txorigin);
	receiver.transfer(amount);
}
\end{lstlisting}
    \caption{Simple source code use \texttt{tx.origin}}
    \label{fig:tx_origin}
\end{figure}{}

\begin{figure}[H]
    \centering
    \lstset{
        lineskip={1.85pt}
    }
\begin{lstlisting}[language=Java]
{
    address tmpVar = tx.origin;
    require (tmpVar == owner_txorigin);
    receiver.transfer(amount);
}
\end{lstlisting}
    \caption{Simple source code use \texttt{tx.origin} after transform}
    \label{fig:tx_origin_m}
\end{figure}{}

Fig.~\ref{fig:tx_origin_m} illustrates the transformed code derived from the original code presented in Fig.~\ref{fig:tx_origin} that show \texttt{tx.origin} is replaced by \texttt{tmpVar}. This intermediary variable is subsequently employed in potentially vulnerable function calls, such as \texttt{require}.

\section{Experiments}
\subsection{Experiment Setup}
For the purposes of this evaluation, a subset of analysis tools was chosen from the original SolidiFI experiment setup. Specifically, we selected two prominent tools: \textit{Slither}, which leverages static analysis techniques, and \textit{Mythril}, which employs symbolic execution. The decision to exclude other tools was due to their discontinued status. Additionally, a state-of-the-art research approach, \textit{CrossFuzz}, was incorporated to explore the effectiveness of fuzzing techniques. All tools were tested with their default configurations.

The original dataset is 272 Solidity code snippets from 7 vulnerability types: Overflow-underflow, Reentrancy, TOD, Timestamp dependency, Unchecked send, Unhandled exceptions and \texttt{tx.origin}.
\subsection{Metric}
A comprehensive evaluation of the generated dataset is crucial to ensure its effectiveness in facilitating research on smart contract vulnerability detection. This evaluation should encompass both quantitative and qualitative metrics. Quantitatively, we assess the volume of transformed bug snippets and the number of newly introduced vulnerable locations. Qualitatively, we examine the ratio of false negative reports.

Formular~\ref{equa:ratio} represents the ratio of the number of false negative cases reported by the tool to the total number of bugs found in the dataset \texttt{D}. If this ratio on the generated dataset is higher than the original dataset, this means the newly created vulnerabilities are complex enough to bypass analysis tools. In this formula,
\begin{itemize}
    \item $n_{FN}$ is the number of false negative cases reported by the tool. 
    \item $N_{D}$ is the number of injected cases to all smart contracts by dataset $D$.
\end{itemize}

\begin{equation}
\label{equa:ratio}
r_{D} = \frac{n_{FN}}{N_{D}}
\end{equation}

\subsection{Experiment result}

\begin{sidewaystable}
\def\arraystretch{1.5}
\centering
\caption{Experiment result}
\begin{tabular}{|c|cccc|ccc|ccc|ccc|cc|cc|}
    \hline
    \multirow{2}{*}{Operator} & \multicolumn{4}{c|}{Reentrancy} & \multicolumn{3}{c|}{Timestamp-Dependency} & \multicolumn{3}{c|}{Overflow-Underflow} & \multicolumn{3}{c|}{tx.origin} & \multicolumn{2}{c|}{Unchecked send} & \multicolumn{2}{c|}{TOD} \\ \cline{2-18} 
     & \multicolumn{1}{c|}{\rotatebox[origin=c]{270}{Vulnerable location}} & \multicolumn{1}{c|}{\rotatebox[origin=c]{270}{Slither}} & \multicolumn{1}{c|}{\rotatebox[origin=c]{270}{Mythril}} & \rotatebox[origin=c]{270}{CrossFuzz} & \multicolumn{1}{c|}{\rotatebox[origin=c]{270}{Vulnerable location}} & \multicolumn{1}{c|}{\rotatebox[origin=c]{270}{Slither}} & \rotatebox[origin=c]{270}{Mythril} & \multicolumn{1}{c|}{\rotatebox[origin=c]{270}{Vulnerable location}} & \multicolumn{1}{c|}{\rotatebox[origin=c]{270}{Mythril}} & \rotatebox[origin=c]{270}{CrossFuzz} & \multicolumn{1}{c|}{\rotatebox[origin=c]{270}{Vulnerable location}} & \multicolumn{1}{c|}{\rotatebox[origin=c]{270}{Slither}} & \rotatebox[origin=c]{270}{Mythril} & \multicolumn{1}{c|}{\rotatebox[origin=c]{270}{Vulnerable location}} & \rotatebox[origin=c]{270}{Mythril} & \multicolumn{1}{c|}{\rotatebox[origin=c]{270}{Vulnerable location}} & \rotatebox[origin=c]{270}{CrossFuzz} \\ \hline
    Default & \multicolumn{1}{c|}{1343} & \multicolumn{1}{c|}{0} & \multicolumn{1}{c|}{1122} & 1047 & \multicolumn{1}{c|}{1381} & \multicolumn{1}{c|}{251} & 586 & \multicolumn{1}{c|}{1333} & \multicolumn{1}{c|}{1333} & 899 & \multicolumn{1}{c|}{1336} & \multicolumn{1}{c|}{2} & 1336 & \multicolumn{1}{c|}{1266} & 968 & \multicolumn{1}{c|}{1336} & 1336 \\ \hline
    $Op_{rename}$ & \multicolumn{1}{c|}{1346} & \multicolumn{1}{c|}{0} & \multicolumn{1}{c|}{1186} & \textbf{1148} & \multicolumn{1}{c|}{1381} & \multicolumn{1}{c|}{251} & \textbf{602} & \multicolumn{1}{c|}{1336} & \multicolumn{1}{c|}{1336} & 846 & \multicolumn{1}{c|}{1369} & \multicolumn{1}{c|}{\textbf{210}} & 1369 & \multicolumn{1}{c|}{1266} & 970 & \multicolumn{1}{c|}{1336} & 1336 \\ \hline
    Permutation & \multicolumn{1}{c|}{-} & \multicolumn{1}{c|}{-} & \multicolumn{1}{c|}{-} & - & \multicolumn{1}{c|}{-} & \multicolumn{1}{c|}{-} & - & \multicolumn{1}{c|}{495} & \multicolumn{1}{c|}{495} & \textbf{495} & \multicolumn{1}{c|}{-} & \multicolumn{1}{c|}{-} & - & \multicolumn{1}{c|}{-} & - & \multicolumn{1}{c|}{-} & - \\ \hline
    $Op_{if2}$ & \multicolumn{1}{c|}{1235} & \multicolumn{1}{c|}{\textbf{131}} & \multicolumn{1}{c|}{1031} & 930 & \multicolumn{1}{c|}{1235} & \multicolumn{1}{c|}{0} & 324 & \multicolumn{1}{c|}{-} & \multicolumn{1}{c|}{-} & - & \multicolumn{1}{c|}{-} & \multicolumn{1}{c|}{-} & - & \multicolumn{1}{c|}{-} & - & \multicolumn{1}{c|}{938} & 937 \\ \hline
    $Op_{if2for}$ & \multicolumn{1}{c|}{1235} & \multicolumn{1}{c|}{\textbf{2}} & \multicolumn{1}{c|}{1069} & 939 & \multicolumn{1}{c|}{1235} & \multicolumn{1}{c|}{0} & 324 & \multicolumn{1}{c|}{-} & \multicolumn{1}{c|}{-} & - & \multicolumn{1}{c|}{-} & \multicolumn{1}{c|}{-} & - & \multicolumn{1}{c|}{-} & - & \multicolumn{1}{c|}{938} & 938 \\ \hline
    $Op_{if2while}$ & \multicolumn{1}{c|}{1235} & \multicolumn{1}{c|}{0} & \multicolumn{1}{c|}{1079} & 964 & \multicolumn{1}{c|}{1235} & \multicolumn{1}{c|}{0} & 324 & \multicolumn{1}{c|}{-} & \multicolumn{1}{c|}{-} & - & \multicolumn{1}{c|}{-} & \multicolumn{1}{c|}{-} & - & \multicolumn{1}{c|}{-} & - & \multicolumn{1}{c|}{938} & 936 \\ \hline
    $Op_{tx}$ & \multicolumn{1}{c|}{1078} & \multicolumn{1}{c|}{0} & \multicolumn{1}{c|}{1001} & 1004 & \multicolumn{1}{c|}{495} & \multicolumn{1}{c|}{0} & 144 & \multicolumn{1}{c|}{938} & \multicolumn{1}{c|}{938} & 345 & \multicolumn{1}{c|}{1336} & \multicolumn{1}{c|}{\textbf{1336}} & 1336 & \multicolumn{1}{c|}{-} & - & \multicolumn{1}{c|}{938} & 938 \\ \hline
\end{tabular}
\label{tab:result}
\end{sidewaystable}

To evaluate the quantity, the total number of new code snippets containing vulnerabilities has grown to \textbf{674}, representing a \textbf{248\%} increase from the original 272. Initially, in terms of goals we see a clear improvement in the approach of creating variations using semantic-preserving code transformations.

Additionally, if we combine the combinatorial transformations, except that $Op_{if2for}$ and $Op_{if2while}$ are not used together, we have a total up to $\sum_{i=1}^{7} C_{7}^{i} - \sum_{i=0}^{5} C_{5}^{i} = \textbf{95}$ operators. With this amount of data, we can use it for machine learning problems, especially new vulnerability detection methods to achieve greater efficiency.

Table~\ref{tab:result} summarizes the experimental outcomes for all operators applied to each analysis tool. The results reveal most transformations significantly elevate the false negative rate $r$ and can up to \textbf{100\%} on Reentrancy, reported by Slither. Notably, certain vulnerabilities are entirely detectable in the original dataset. For instance, Slither successfully identifies all vulnerable locations associated with Reentrancy and \texttt{tx.origin}. However, after applying the transformation operators, Slither exhibits a substantial increase in false negatives. With the new dataset, we can detect more cases that analytics tools cannot identify and improve. An increase in the $r$ ratio appears for all 3 tools in the experiment.

This experimental result shows that having solved the first research question (RQ1), the number of vulnerable increased significantly, and at the same time, they are more difficult to detect by current tools.

Permutation is shown to influence computations, leading to incorrect identification of underflow or overflow computations. Using an intermediate variable in $Op_{tx}$ makes it difficult for tools to detect errors using the \texttt{tx.origin} expression. Statement transformations affect vulnerabilities related to multiple statements such as Reentrancy. Meanwhile, renaming does not have much effect, but in some cases, it is possible to inject some additional locations by renaming so that the snippets do not conflict with each other's function names.

We extracted random cases with false negative results and tested stand-alone this case. We consider making sure that there are no mistakes during testing. These cases that have code snippets after transform can bypass the detector of the above tools, while their original code snippet cannot. Notably, we observed that certain code transformations, such as statement reversal and the addition of negation symbols, could render the code undetectable by the tools, while the original code snippet was flagged appropriately. This is a limitation in the detection capabilities of these tools for specific code transform techniques.

\section{Related Work}
SolidiFI injects vulnerabilities at all potential locations, instead of a single vulnerability \cite{ghaleb2020effective}. They explain the method can find more corner cases that cannot be easily detected. However, the vulnerabilities injected into their dataset are independent, so they will have the same semantics in different locations and they still retain the characteristic of being vulnerable. In essence, a vulnerability at any location has the same semantics. Nevertheless, this methodology enables us to more precisely identify and assess the limitations of the analyzer.

To verify our findings and answer RQ2, we conducted a small experiment. We try to inject all Reentrancy vulnerabilities transformed by the $Op_{if}$ operator and evaluate the ratio $r$ of Slither compared to injecting at all potential locations. The number of locations of each snippet is the same.

Table~\ref{tab:onelocation} shows that injecting vulnerability to only one position produces 10\% of false negative rate. Meanwhile, injecting to all potential positions reaches 10.6\%, presented in Table~\ref{tab:result}. This suggests that Slither produced more false positives in contracts with multiple potential injection locations.

\begin{table}[H]
\def\arraystretch{1.25}
\caption{Slither result on transformed Reentrancy dataset}
\centering
\begin{tabular}{|c|c|c|}
\hline
Vulnerability type & Vulnerable location & False negative \\ \hline
Reentrancy        & 30                  & 3              \\ \hline
\end{tabular}
\label{tab:onelocation}
\end{table}

Indeed, as Slither reports, there are some cases where the detector of this tool doesn't cover loops or abstract/library syntax\footnote{https://github.com/crytic/slither/pull/2419}. Such bugs remain unidentified until user intervention. The lack of high-quality datasets makes it hard for tools to test their issues automatically. In this context, the proposed injection method targeting all potential insertion locations serves a two-fold purpose: facilitating the assessment and accurate evaluation of analysis tools and concurrently uncovering errors within the tools themselves, particularly those that might be easily missed.

\section{Conclusions}
Reliable data is essential for research in software vulnerability detection problems. Unfortunately, the domain of smart contract security faces a significant hurdle due to the scarcity of such data. This paper proposes a novel method specifically designed to generate datasets tailored to smart contract vulnerabilities. This method leverages semantic-preserving code transformation techniques, demonstrably enhancing both the quantity and quality of existing datasets. The results show that the increased amount of data simultaneously increases the false negative rate on analysis tools. This also shows that the detection tools are sensitive to source code variants that have undergone transform operators.

These enriched datasets will pave the way for significant advancements in smart contract vulnerability detection accuracy. From there, it is possible to develop new detection methods such as machine learning and deep learning in this problem. Consequently, the potential for successful attacks on high-value blockchain systems can be substantially reduced.

\section*{Acknowledgment}
This work has been supported by VNU University of Engineering and Technology under project number CN24.12.

% \section*{References}

\printbibliography

@article{yaga2019blockchain,
  title={Blockchain technology overview},
  author={Yaga, Dylan and Mell, Peter and Roby, Nik and Scarfone, Karen},
  journal={arXiv preprint arXiv:1906.11078},
  year={2019}
}

@inproceedings{dhawan2017analyzing,
  title={Analyzing safety of smart contracts},
  author={Dhawan, Mohan},
  booktitle={Proceedings of the Conference: network and distributed system security symposium, San Diego, CA, USA},
  pages={16--17},
  year={2017}
}

@article{liu2021combining,
  title={Combining graph neural networks with expert knowledge for smart contract vulnerability detection},
  author={Liu, Zhenguang and Qian, Peng and Wang, Xiaoyang and Zhuang, Yuan and Qiu, Lin and Wang, Xun},
  journal={IEEE Transactions on Knowledge and Data Engineering},
  volume={35},
  number={2},
  pages={1296--1310},
  year={2021},
  publisher={IEEE}
}

@article{sun2023assbert,
  title={ASSBert: Active and semi-supervised bert for smart contract vulnerability detection},
  author={Sun, Xiaobing and Tu, Liangqiong and Zhang, Jiale and Cai, Jie and Li, Bin and Wang, Yu},
  journal={Journal of Information Security and Applications},
  volume={73},
  pages={103423},
  year={2023},
  publisher={Elsevier}
}

@article{zhen2024gnn,
  title={DA-GNN: A smart contract vulnerability detection method based on Dual Attention Graph Neural Network},
  author={Zhen, Zixian and Zhao, Xiangfu and Zhang, Jinkai and Wang, Yichen and Chen, Haiyue},
  journal={Computer Networks},
  pages={110238},
  year={2024},
  publisher={Elsevier}
}

@inproceedings{mi2021vscl,
  title={VSCL: automating vulnerability detection in smart contracts with deep learning},
  author={Mi, Feng and Wang, Zhuoyi and Zhao, Chen and Guo, Jinghui and Ahmed, Fawaz and Khan, Latifur},
  booktitle={2021 IEEE International Conference on Blockchain and Cryptocurrency (ICBC)},
  pages={1--9},
  year={2021},
  organization={IEEE}
}

@inproceedings{kien2023multimodal,
  title={A Multimodal Deep Learning Approach for Efficient Vulnerability Detection in Smart Contracts},
  author={Kien, Vu Trung and Hoang, Trinh Minh and Quyen, Nguyen Huu and Khoa, Nghi Hoang and Duy, Phan The and Pham, Van-Hau and others},
  booktitle={GLOBECOM 2023-2023 IEEE Global Communications Conference},
  pages={3421--3426},
  year={2023},
  organization={IEEE}
}

@inproceedings{ghaleb2020effective,
  title={How effective are smart contract analysis tools? evaluating smart contract static analysis tools using bug injection},
  author={Ghaleb, Asem and Pattabiraman, Karthik},
  booktitle={Proceedings of the 29th ACM SIGSOFT International Symposium on Software Testing and Analysis},
  pages={415--427},
  year={2020}
}

@article{chu2023survey,
  title={A survey on smart contract vulnerabilities: Data sources, detection and repair},
  author={Chu, Hanting and Zhang, Pengcheng and Dong, Hai and Xiao, Yan and Ji, Shunhui and Li, Wenrui},
  journal={Information and Software Technology},
  pages={107221},
  year={2023},
  publisher={Elsevier}
}

@inproceedings{liao2019soliaudit,
  title={Soliaudit: Smart contract vulnerability assessment based on machine learning and fuzz testing},
  author={Liao, Jian-Wei and Tsai, Tsung-Ta and He, Chia-Kang and Tien, Chin-Wei},
  booktitle={2019 Sixth International Conference on Internet of Things: Systems, Management and Security (IOTSMS)},
  pages={458--465},
  year={2019},
  organization={IEEE}
}

@inproceedings{ghaleb2022towards,
  title={Towards effective static analysis approaches for security vulnerabilities in smart contracts},
  author={Ghaleb, Asem},
  booktitle={Proceedings of the 37th IEEE/ACM International Conference on Automated Software Engineering},
  pages={1--5},
  year={2022}
}

@article{colin2024integrated,
  title={An Integrated Smart Contract Vulnerability Detection Tool Using Multi-layer Perceptron on Real-time Solidity Smart Contracts},
  author={Colin, Lee Song Haw and Mohan, Purnima Murali and Pan, Jonathan and Keong, Peter Loh Kok},
  journal={IEEE Access},
  year={2024},
  publisher={IEEE}
}

@article{zhang2023challenging,
  title={Challenging machine learning-based clone detectors via semantic-preserving code transformations},
  author={Zhang, Weiwei and Guo, Shengjian and Zhang, Hongyu and Sui, Yulei and Xue, Yinxing and Xu, Yun},
  journal={IEEE Transactions on Software Engineering},
  year={2023},
  publisher={IEEE}
}

@article{le2024evaluating,
  title={Evaluating Program Repair with Semantic-Preserving Transformations: A Naturalness Assessment},
  author={Le-Cong, Thanh and Nguyen, Dat and Le, Bach and Murray, Toby},
  journal={arXiv arXiv:2402.11892},
  year={2024}
}

@inproceedings{kado2023empirical,
  title={An empirical study of impact of solidity compiler updates on vulnerabilities},
  author={Kado, Chihiro and Yanai, Naoto and Cruz, Jason Paul and Okamura, Shingo},
  booktitle={2023 IEEE International Conference on Pervasive Computing and Communications Workshops and other Affiliated Events (PerCom Workshops)},
  pages={92--97},
  year={2023},
  organization={IEEE}
}

\end{document}